\documentclass[twocolumn,aps,pre,floatfix,superscriptaddress,longbibliography]{revtex4-1}
\pdfoutput=1 
\usepackage{amsmath,amssymb,eucal,graphicx,float,epstopdf}

\usepackage[colorlinks=true, urlcolor=blue, anchorcolor=blue, citecolor=blue,filecolor=blue,linkcolor=blue,menucolor=blue]{hyperref}

\begin{document}
\title{Templating Aggregation}
\author{P. L. Krapivsky}
\affiliation{Department of Physics, Boston University, Boston, MA 02215, USA}
\affiliation{Santa Fe Institute, 1399 Hyde Park Road, Santa Fe, NM 87501, USA}
\author{S. Redner}
\affiliation{Santa Fe Institute, 1399 Hyde Park Road, Santa Fe, NM 87501, USA}

\begin{abstract}

  We introduce an aggregation process based on \emph{templating},
  where a specified number of constituent clusters must assemble on a
  larger aggregate, which serves as a scaffold, for a reaction to
  occur.  A simple example is a dimer scaffold, upon which two
  monomers meet and create another dimer, while dimers and larger
  aggregates undergo in irreversible aggregation with mass-independent
  rates.  In the mean-field approximation, templating aggregation has
  unusual kinetics in which the cluster and monomer densities, $c(t)$
  and $m(t)$ respectively, decay with time as
  $c\sim m^2\sim t^{-2/3}$.  These starkly contrast to the
  corresponding behaviors in conventional aggregation,
  $c\sim \sqrt{m}\sim t^{-1}$. We then treat three natural extensions
  of templating: (a) the reaction in which $L$ monomers meet and react
  on an $L$-mer scaffold to create two $L$-mers, (b) multistage
  scaffold reactions, and (c) templated ligation, in which clusters of
  all masses serve as scaffolds and binary aggregation is absent.
  
\end{abstract}  
\maketitle

\section{Introduction}

Irreversible aggregation is a fundamental kinetic process in which two
clusters from a heterogeneous population irreversibly merge to form a
larger cluster.  We may represent the reaction as
\begin{align}
\label{agg}
C_i\oplus C_j\mathop{\longrightarrow}^{K_{i,j}}C_{i+j}\,,
\end{align}
where $C_i$ denotes a cluster of mass $i$ and $K_{i,j}$ specifies the
rate at which a cluster of mass $i$ (an $i$-mer) joins to a $j$-mer to
form an $(i+j)$-mer.  The basic observable in aggregation is the
cluster-mass distribution, whose properties depend on the functional
form of the reaction kernel $K_{i,j}$.  In the mean-field
approximation in which all reactants are perfectly mixed, the time
dependence of the cluster-mass distribution is described by an
infinite set of rate equations that accounts for the change in the
cluster concentrations due to reactions with other clusters.

\begin{figure}[ht]
\includegraphics[width=0.48\textwidth]{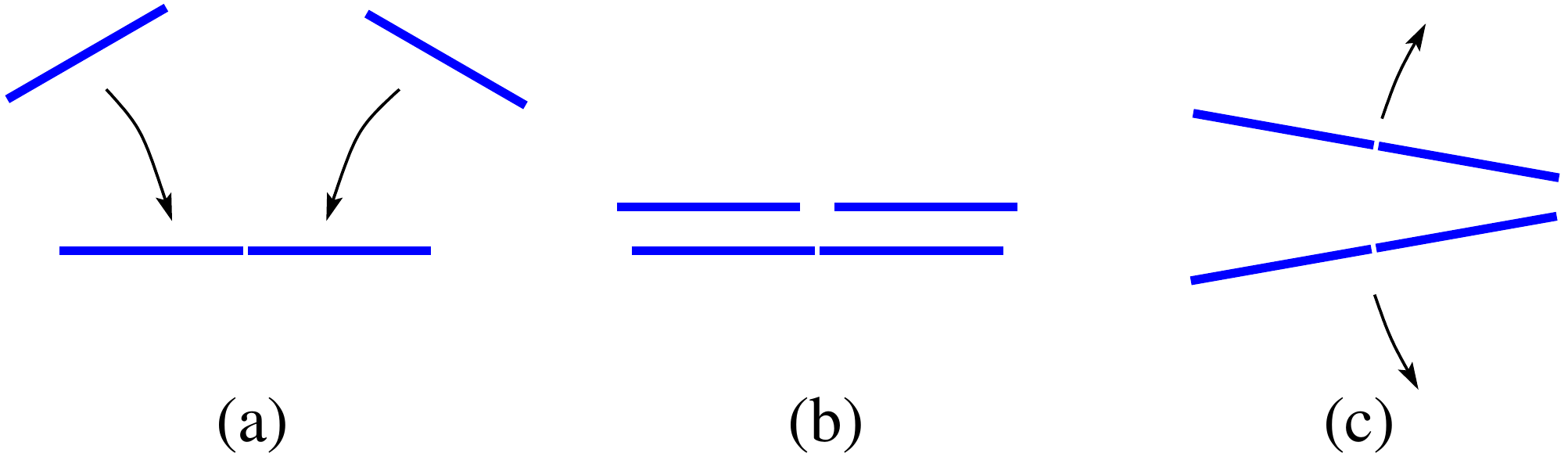}
\caption{The steps in templating aggregation on a dimer scaffold: (a)
  two monomers simultaneously meet on the scaffold and (b) form a
  second dimer, so that (c) two dimers result.}
\label{fig:model}
\end{figure}

The emergence of complex molecules from prebiotic building blocks is a
key aspect in theories of the origin of
life~\cite{eigen1971selforganization,kauffman1971cellular,gilbert1986origin,Nowak08,Nowak09,Nigel15,Kaneko16,hordijk2019history}.
Aggregation processes that generate growing (and hence more complex)
clusters provide a convenient starting point for theoretical analyses.
Pure aggregation is too minimal a process, and one would like to
enrich the reaction scheme \eqref{agg} by additional processes that
contribute to the emergence of complex entities. In this work, we
investigate an aggregation process that is augmented by the mechanism
of \emph{templating}.  Here a cluster of a specified mass $s$ serves
as a scaffold that facilitates the reaction (Fig.~\ref{fig:model}).
On this scaffold, two clusters of masses $t<s$ and $s-t$ meet and
merge to form another cluster of mass $s$.  Clusters of mass $s$ can
continue to serve as scaffolds for subsequent reactions or they can
participate in binary aggregation.  The templating reaction can be
viewed as the autocatalytic replication of scaffolds, a reaction step
that seems to be an essential feature in various origin of life
models~\cite{Steen14,Steen17,Kaneko18,tkachenko18,Kaneko23,SA21,Kaneko23}; this
is the underlying motivation for our model.

Our goal is to determine the kinetics of this template-controlled
aggregation.  We first treat a simple version of templating
aggregation in which the scaffolds are dimers and all reaction rates
are mass independent. When two monomers meet on a dimer, the latter
serves as a scaffold, and the monomers merge to create another dimer.
Symbolically, this template-controlled merging of dimers is
represented by the reaction scheme
\begin{align}
\label{MM}
  M\oplus M\oplus D\to D+D\,,
\end{align}
where $M$ denotes a monomer and $D$ denotes a dimer.  Clusters of
masses greater than or equal to 2 (the dimer mass), undergo
conventional binary aggregation.  Thus, the overall reaction is
comprised of 2-body and 3-body processes.  This mixture of different
reaction orders underlies the unusual kinetics of our model.

For binary aggregation with mass independent reaction rates and the
monomer-only initial condition, the density of clusters of mass $k$ at
time $t$, $c_k(t)$, is given
by~\cite{Smo17,Drake,Chandra,Leyvraz03,KRB}
\begin{align*}
  \label{ck-classic}
  c_k(t) = \frac{1}{(1+t)^2}\left(\frac{t}{1+t}\right)^{k-1}
  \mathop{\longrightarrow}_{t\to\infty}~~ \frac{1}{t^2}\; e^{-k/t}\,.
\end{align*}
From this solution, the monomer density $c_1(t)$ and the total cluster
density, $c(t) \equiv \sum_{k\geq 1} c_k(t)$ both decay algebraically
with time:
\begin{equation}
\label{c-c1}
c_1\simeq \frac{1}{t^{2}}\,, \qquad c\simeq \frac{1}{t}\,.
\end{equation}
These decay laws are independent of the initial condition and hence
universal.  For templating aggregation with scaffold mass $s=2$, it is
convenient to denote the monomer density as $m(t)$ and the density of
clusters of mass $2k$ as $c_k(t)$.  In contrast to \eqref{c-c1}, here
we find
\begin{equation}
\label{m-c-c1}
m \simeq \frac{2}{(3t)^{1/3}}\,, \quad 
c_1 \simeq  \frac{1}{2(3t)^{2/3}}\,, \quad 
c \simeq  \frac{1}{(3t)^{2/3}}~.
\end{equation}
Surprisingly, the monomer density decays slower than the cluster
density, while the opposite occurs in conventional aggregation.
Another unusual feature of templating aggregation is that the decay
exponents for dimers and the total cluster density are the same.

In Sec.~\ref{sec:dimer}, we analyze the template-controlled
aggregation with dimer scaffolds and derive the decay laws
\eqref{m-c-c1}, as well as the decay law for $c_k(t)$, the density of
clusters of mass $2k$.  In Sec.~\ref{sec:general}, we study more
general models of template-controlled aggregation. First, we consider
the model with scaffolds of fixed mass $L$ for arbitrary $L\geq
2$. Then we analyze the effect of multiple levels of
templating. Specifically, we treat a model with two types of
templates, dimers and 4-mers,
and the template-controlled reaction in \eqref{MM} is supplemented by the reaction
\begin{align}
\label{DD}
  D\oplus D\oplus F\to F+F\,,
\end{align}
where $F$ represents 4-mers.  Under the assumption that clusters of
mass 4 and greater undergo ordinary aggregation, we find kinetic
behaviors similar to that quoted in Eq.~\eqref{m-c-c1}.  Finally, in
Sec.~\ref{sec:lig}, we study templated ligation, in which clusters of
all masses serve as scaffolds and in which no binary aggregation
reactions occur.  In distinction to the reactions where the scaffolds
have a specified mass, we now find that the cluster density is the
most slowly decaying quantity at long times, with $c\sim t^{-1/3}$,
while the monomer density asymptotically decays $c_1\sim t^{-2/3}$.

\section{Templating with Dimer Scaffolds}
\label{sec:dimer}

Under the assumption that clusters are perfectly mixed and that the
reaction rates are mass-independent, the rate equations for the
monomer and dimer densities are
\begin{align}
\label{template}
\begin{split}
\frac{dm}{dt} &=-m^2 c_1\,,\\
\frac{dc_1}{dt} &=\tfrac{1}{2}m^2 c_1-2cc_1\,,
\end{split}
\end{align}
while the densities of heavier clusters satisfy
\begin{equation}
\label{ck-eq}
\frac{dc_k}{dt} =\sum_{i+j=k}c_i c_j-2c c_k \qquad (k\geq 2)\,.
\end{equation}

The first of Eqs.~\eqref{template} accounts for the decay of monomers
when two monomers meet on a scaffold to create a dimer. Dimer creation
is encapsulated by the first term on the right of the second of
Eqs.~\eqref{template}.  Equations \eqref{ck-eq} account for the
conventional aggregation reactions of heavier-mass clusters.  A useful
check of the consistency of Eqs.~\eqref{template}--\eqref{ck-eq} is to
verify that the mass density
\begin{equation*}
m+2\sum_{k\geq 1}kc_k
\end{equation*}
is conserved.  For templating aggregation, we must additionally
postulate that both $m(0)>0$ and $c_1(0)>0$.  This condition ensures
that scaffolds are always present to catalyze continuous evolution of
the cluster mass distribution.

To determine the asymptotic behavior of templating aggregation, we
begin by adding the second of Eqs.~\eqref{template} and all
Eqs.~\eqref{ck-eq} to give
\begin{equation}
\label{c-eq}
\frac{dc}{dt} =\tfrac{1}{2}m^2 c_1-c^2\,,
\end{equation}
where $c\equiv \sum_{k\geq 1}c_k$ is the total cluster density.  The
first two of Eqs.~\eqref{template} and Eq.~\eqref{c-eq} constitute a
closed system of three differential equations whose solution would
determine the resulting kinetics.

The form of the equation for the monomer density suggests introducing
the modified time variable
\begin{equation}
\label{tau-eq}
\tau = \int_0^t dt'\, c_1(t')\,,
\end{equation}
to recast the rate equation for the monomer density into
$\frac{dm}{d\tau} =-m^2$, with solution
\begin{equation}
\label{m-tau-sol}
m(\tau) = \frac{m(0)}{1+m(0)\tau}\,.
\end{equation}
Since the long-time asymptotic behavior is $m\simeq \tau^{-1}$,
independent of $m(0)$, we adopt the simple initial condition $m(0)=1$
for simplicity henceforth.

Using the time modified time variable \eqref{tau-eq}, as well as the
solution \eqref{m-tau-sol}, we rewrite the equation for the dimer and
cluster concentrations as \eqref{c-eq} as
\begin{align}
\begin{split}
\label{c-c1-tau}
\frac{dc_1}{d\tau} &=\frac{1}{2(1+\tau)^2}-2c\,,\\
\frac{dc}{d\tau}  &= \frac{1}{2(1+\tau)^2}-\frac{c^2}{c_1}\,.
\end{split}
\end{align}
Without loss of generality, we choose the initial condition
$c_1(0)=c(0)=\rho$.  While the full initial-value problem
\eqref{c-c1-tau} subject to this initial condition appears to be
intractable, we can deduce the physically relevant long-time behavior
by the method of dominant balance~\cite{Bender}, in which we neglect
one of the three terms in each of Eqs.~\eqref{c-c1-tau} and check that
the assumption is self-consistent.

By this approach, we deduce that in both Eqs.~\eqref{c-c1-tau}, the
right-hand side (RHS) dominates the left-hand side (LHS). Neglecting
the LHS in Eqs.~\eqref{c-c1-tau}, we find
\begin{equation}
\label{c-c1-asymp}
c\simeq \frac{1}{4(1+\tau)^2}\,, \qquad c_1\simeq \frac{1}{8(1+\tau)^2}\,.
\end{equation}
as $\tau\to\infty$. A more detailed but straightforward asymptotic
analysis of Eqs.~\eqref{c-c1-tau} gives the more complete long-time
behavior
\begin{align}
\begin{split}
\label{complete}
&c_1 = \frac{1}{8(1+\tau)^2} - \frac{1}{16(1+\tau)^4}-\frac{1}{16(1+\tau)^5}\,,\\
&c = \frac{1}{4(1+\tau)^2} + \frac{1}{8(1+\tau)^3} - \frac{1}{8(1+\tau)^5}\,,
\end{split}
\end{align}
where we drop terms of $O(\tau^{-6})$ and lower.  All algebraic
correction terms are universal, i.e., independent of the initial
conditions. Only terms that are exponentially small in the
$\tau\to\infty$ limit depend on the initial condition.

To find the dependence of the cluster densities on the physical time,
we substitute $c_1$ from \eqref{c-c1-asymp} in the definition of the
modified time \eqref{tau-eq} and invert this relation to give
\begin{equation}
\label{t-tau}
t = \int_0^\tau \frac{d\tau'}{c_1(\tau')} \simeq  8 \int_0^\tau d\tau'\,(1+\tau')^2 \simeq \frac{8}{3}\,\tau^3\,.
\end{equation}
If we employ the more accurate asymptotic formula in \eqref{complete}
for $c_1(\tau)$, we instead obtain the expansion
\begin{equation}
\label{t-tau-asymp}
t = \frac{8}{3}\,\tau^3+ 8\tau^2+\frac{11}{6}\,\tau + O(1)\,,
\end{equation}
with three exact terms. The last term, a constant, cannot be
determined analytically since it depends on the initial condition.
Limiting ourselves to the leading asymptotic behavior, we substitute
$\tau\simeq \left({3t}/{8}\right)^{1/3}$ into \eqref{m-tau-sol} and
\eqref{c-c1-asymp} to arrive at the central results given in
Eq.~\eqref{m-c-c1}.  

\begin{figure}[h]
    \centering
\includegraphics[width=0.48\textwidth]{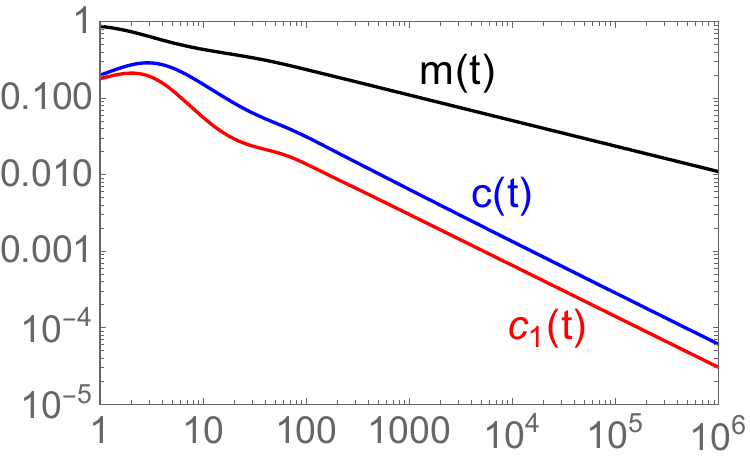}
\caption{Time dependence of $m(t), c(t)$, and $c_1(t)$ on a double
  logarithmic scale, with asymptotic decay of $t^{-1/3}$ for $m(t)$
  and $t^{-2/3}$ for both $c(t)$ and $c_1(t)$, as predicted by
  Eq.~\eqref{m-c-c1}.  The ratio $c(t)/c_1(t)$ quickly approaches 2
  for increasing time, as predicted by \eqref{c-c1-asymp}.}
\label{fig:m-c-c1}
\end{figure}

Figure~\ref{fig:m-c-c1} shows the time dependence of $m(t)$, $c(t)$
and $c_1(t)$ obtained by numerical integration of the first two of
Eqs.~\eqref{template} and Eq.~\eqref{c-eq} by Mathematica. As the
initial condition we use $m(0)=1$ and $c_1(0)=c(0)=0.1$.

Having found the dimer concentration $c_1(t)$, we now outline how to
obtain all the cluster densities.  Based on the form of the rate
equation for $c_k(t)$ for $k\geq 2$, we anticipate that $c_k=A_k c$;
that is, all individual cluster densities are of the same order as the
total cluster density.  Substituting this ansatz into the last of
Eqs.~\eqref{template} we obtain the following recursion for the
amplitudes:
\begin{align}
\label{Ak-eq}
  \sum_{i+j=k}A_i A_j=2A_k\qquad  k\geq 2\,.
\end{align}
The already known value $A_1=\frac{1}{2}$ plays the role of the
initial condition for this recursion.  Solving \eqref{Ak-eq} subject
to this initial condition gives
\begin{equation}
\label{ratios}
\frac{c_k}{c} = \frac{1}{\sqrt{4\pi}}\,\frac{\Gamma\big(k-\frac{1}{2}\big)}{\Gamma(k+1)}
\simeq  \frac{1}{\sqrt{4\pi}}\, \frac{1}{k^{3/2}}\,,
\end{equation}
where the last asymptotic is valid when $k\gg 1$ and generally \eqref{ratios} holds 
in the $t\to\infty$  limit.

The large-$k$ asymptotic, $A_k\sim k^{-3/2}$, holds only up to a
cutoff value $k^*$, beyond which the $A_k$ must decay faster than any
power law.  To understand the origin of this cutoff, we note that the
sum $\sum_{k\geq 1} kA_k$ diverges if $A_k\sim k^{-3/2}$ for all $k$.
This divergence contradicts the mass conservation statement that
$\sum_{k\geq 1} kc_k \to \text{const}$.  To resolve this apparent
divergence, the power-law behavior of $A_k$ must break down beyond a
cutoff value $k^*$.  To determine $k^*$, we compute
\begin{align*}
\sum_{k\geq 1} kc_k \sim c \sum_{1\leq k\leq k^*}k\times k^{-3/2}\sim
  c\sqrt{k^*}\,.
\end{align*}
Since both sums are constant, we see that the threshold mass is given
by $k^*\sim c^{-2}\sim \tau^4\sim t^{4/3}$.  Thus the power-law mass
distribution is cut off at $k^*$ to ensure mass conservation.  This
cutoff is analogous to what happens in constant-kernel aggregation
with a steady monomer source~\cite{KRB}.  In this latter example, the
cutoff is determined by the condition that the total mass in the
system equals to the total mass that is injected up to a given time.

\section{More General Templating Reactions}
\label{sec:general}

There are two natural generalizations of templating aggregation that
we now explore.  One generalization is to consider scaffolds that are
heavier than dimers, and another to analyze what happens when there
are multiple stages of scaffold reactions.

\subsection{Templating with scaffolds of mass $L$}
\label{subsec:L}

Suppose the scaffold has mass $L$ and the simultaneous presence of $L$
monomers on the scaffold is required to create a second $L$-mer.
Aggregates of mass $L$ and heavier also undergo conventional
aggregation. We assume that the process begins with monomers and
scaffolds. By construction, the masses of heavier aggregates are
integer multiples of $L$.

We can determine the kinetics of this model by adapting the analysis
of the previous section in a straightforward way.  In what follows
$c_k$ now denotes the concentration of clusters of mass $kL$.  Within
this model, the first two of Eqs.~\eqref{template} and
Eq.~\eqref{c-eq} become
\begin{align}
\label{template-L}
\begin{split}
\frac{dm}{dt} &=-m^L c_1\,,\\[1mm]
\frac{dc_1}{dt} &=\tfrac{1}{L}\,m^L c_1-2cc_1\,,\\[1mm]
\frac{dc}{dt} &=\tfrac{1}{L}\,m^L c_1-c^2\,.
\end{split}
\end{align}
In terms of the modified time variable defined in \eqref{tau-eq}, the
solution for the monomer density for the initial condition $m(0)=1$
now is
\begin{align}
\label{m-L}
  m(\tau) = \frac{1}{[1+(L-1)\tau]^{1/(L-1)}}\,.
\end{align}
Using this solution for the monomer density and also employing the
same dominant balance method as in the previous section, we obtain
\begin{align}
\label{c-c1-L}
\begin{split}
    c&\simeq \frac{1}{2L} \,\frac{1}{[1+(L-1)\tau]^{L/(L-1)}}~,\\
c_1&\simeq \frac{1}{4L} \,\frac{1}{[1+(L-1)\tau]^{L/(L-1)}}~.
\end{split}
\end{align}

We now express these two densities in terms of the physical time:
\begin{align}
\label{t-tau-L}
t = \int \frac{d\tau'}{c_1(\tau')} \simeq \frac{4L}{2L-1}\; [1+(L-1)\tau]^{(2L-1)/(L-1)}\,.
\end{align}
Combining \eqref{m-L} and \eqref{c-c1-L} with \eqref{t-tau-L}, we
thereby find the densities of monomers, $L-$mers, and the total
cluster density decay as
\begin{align}
\begin{split}
 & m(t) \simeq \left(\frac{2L-1}{4L}\,t\right)^{-1/(2L-1)}\,,\\
 & c(t) \simeq  \frac{1}{2L} \,\left(\frac{2L-1}{4L}\,t\right)^{-L/(2L-1)}\,,\\
& c_1(t) \simeq  \frac{1}{4L} \,\left(\frac{2L-1}{4L}\,t\right)^{-L/2L-1)}\,.
\end{split}
\end{align}
As one might expect, the overall reaction kinetics slows down as the
scaffold size and consequently the reaction order $L$ increases. The
ratios $c_k/c$ are again stationary in the long-time limit and are
given by the same formula \eqref{ratios} as for $L=2$.  Stationarity
again holds up to a threshold mass $k^*$ that grows as
$k^*\sim c^{-2} \sim t^{2L/(2L-1)}$.

\subsection{Multiple levels of templating}
\label{subsec:multi}

\begin{figure}[ht]
  \includegraphics[width=0.48\textwidth]{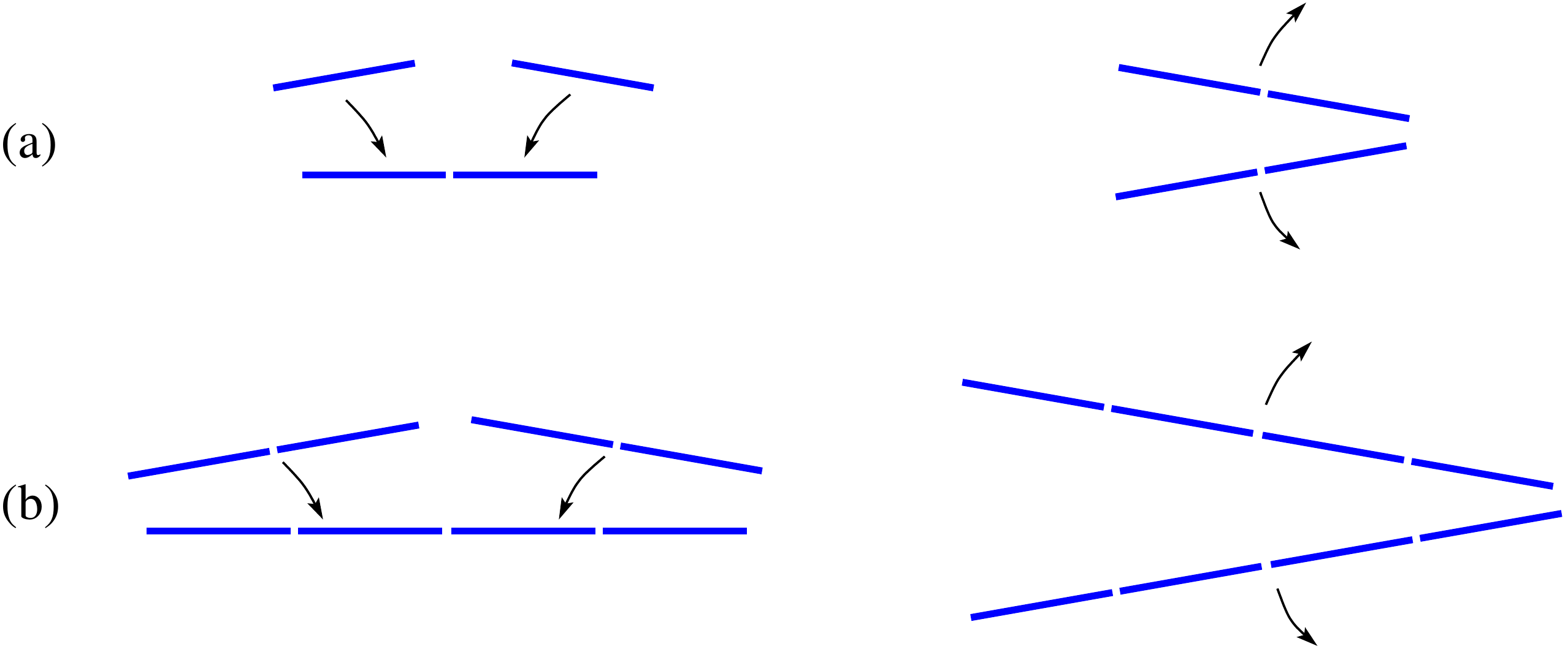}
  \caption{Templating aggregation with two levels of templating: (a)
    two monomers react on a dimer scaffold, (b) two dimers react on a
    4-mer scaffold.}
\label{fig:model4}
\end{figure}

Another natural scenario is a reaction that relies on multiple levels
of templating.  Here we treat the simplest case of two levels of
templating in which: (a) a new dimer template is created when two
monomers react on an existing dimer template, and (b) a new 4-mer
template is created when two dimers react on an existing 4-mer
template (Fig.~\ref{fig:model4}).  In this formulation, dimers are not
free to aggregate; only clusters of mass 4 and greater can react via
conventional aggregation.

Let $m(t)$ and $D(t)$ denote the density of monomers and dimers,
respectively, and let $c_k$ with $k\geq 1$ now denote the density of
clusters of mass $4k$.  In close analogy with Eqs.~\eqref{template},
the rate equations for the various cluster densities now are:
\begin{align}
\label{template-multi}
\begin{split}
\frac{dm}{dt} &=-m^2 D\,,\\[1mm]
\frac{dD}{dt} &=\tfrac{1}{2} m^2 D -D^2 c_1\,,\\[1mm]
\frac{dc_1}{dt} &=\tfrac{1}{2}D^2 c_1-2cc_1\,,
\end{split}
\end{align}
while the densities of heavier clusters satisfy Eqs.~\eqref{ck-eq}.
Summing the last of Eqs.~\eqref{template-multi} and all of
Eqs.~\eqref{ck-eq} we deduce the evolution equation for the total
cluster density
\begin{align}
 \label{c-multi}
  \frac{dc}{dt} = \tfrac{1}{2}D^2 c_1 -c^2\,.
\end{align}

\begin{figure}[h]
    \centering
\includegraphics[width=0.45\textwidth]{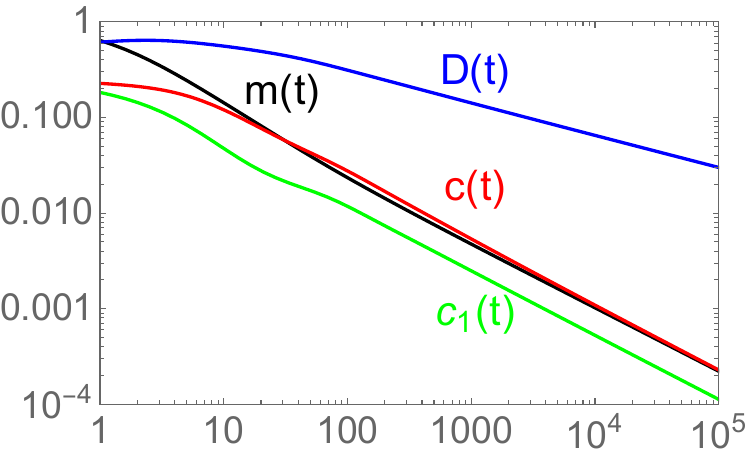}
\caption{Time dependence of $m(t), D(t), c(t)$, and $c_1(t)$ on a
  double logarithmic scale, with asymptotic decay of $t^{-1/3}$ for
  $D(t)$ and $t^{-2/3}$ for $m(t)$, $c(t)$, and $c_1(t)$. }
\label{fig:m-D-c-c1}
\end{figure}

Equations \eqref{template-multi} and \eqref{c-multi} again constitute
a closed system from which we can, in principle, determine the kinetic
behavior.  However, these coupled nonlinear equations do not possess
an exact solution.  Instead, we again use the method of dominant
balance to infer the asymptotic behavior.  We first neglect the LHS in
the last of Eqs.~\eqref{template-multi} to give $c\simeq D^2/4$.  We
substitute this result into \eqref{c-multi}, where we also neglect the
LHS to find $c_1\simeq D^2/8$. Thus
\begin{equation}
\label{cc1:D}
c\simeq \tfrac{1}{4}D^2\,, \quad c_1\simeq \tfrac{1}{8}D^2\,.
\end{equation}

There are various choices of which terms to neglect in
Eqs.~\eqref{template-multi}, and the dominant balance that proves
consistent is to keep the LHS in the second of
Eqs.~\eqref{template-multi}.  Since this term is negative, the
simplest choice now is to neglect the first term on the RHS of this
equation.  We thus have $\frac{dD}{dt} =-D^4/8$. Solving this equation
and substituting this solution into \eqref{cc1:D} and also into the
first of Eqs.~\eqref{template-multi} and solving these two equations
we finally obtain
\begin{equation}
\label{mcc1:D}
m \simeq  c\simeq 2c_1 \simeq (3t)^{-2/3}\,, \quad D \simeq 2 (3t)^{-1/3}\,.
\end{equation}
The decay laws for $c$ and $c_1$ are the same as in the templating
with dimer scaffolds, and even the amplitudes are identical
[cf. \eqref{m-c-c1}]. The density of monomers decays similarly to $c$
and $c_1$, and only the dimer density has the slowest decay of
$t^{-1/3}$.  We used Mathematica to numerically integrate
Eqs.~\eqref{template-multi} and \eqref{c-multi} and the results are
shown in Fig.~\ref{fig:m-D-c-c1}. The asymptotic behaviors are in
excellent agreement with the theoretical predictions \eqref{mcc1:D}.

\section{Templated Ligation}
\label{sec:lig}

We now investigate a self-templating reaction in which clusters of all
masses can serve as scaffolds.  This templated ligation
process~\cite{Steen14,Steen17,MANAPAT2010317,10.1093/nar/gks065,SA21} is
represented by the reaction scheme
\begin{align}
\label{lig}
C_i\oplus C_j \oplus C_{i+j}~ \mathop{\longrightarrow}^{L_{i,j}} ~ C_{i+j}+C_{i+j}\,.
\end{align}
We assume that this ligation reaction is the only dynamical process in
the system.  In particular, binary aggregation of clusters does not
occur in this model. The absence of aggregation reactions means that
an initially compact mass distribution with a maximum mass $J$ remains
compact forever; that is, $c_j(t)=0$ for all $j>J$.  However, for
unbounded initial mass distributions that decay sufficiently rapidly
with mass, the emergent behaviors are universal, that is,
asymptotically independent of the initial condition.  One such example
is the exponential initial mass distribution $c_j(0)=2^{-j-1}$ whose
mass density is normalized to 1:
\begin{equation}
\label{mass}
\sum_{j\geq 1}c_j(t) = 1
\end{equation}

If the ligation rates $L_{i,j}$ are mass independent, the equations
for the evolution of the cluster-mass distribution are
\begin{equation}
\label{ck-lig}
\frac{dc_k}{dt} = c_k\sum_{i+j=k}c_i c_j-2c_k \sum_{j\geq 1}c_j c_{j+k}\,.
\end{equation}
These equations are hierarchical and therefore appear to be
unsolvable.  A mathematically related non-recurrent structure arises
in the rate equations for cluster eating~\cite{redner1987kinetics} and
for combined aggregation-annihilation reactions~\cite{PK93}, albeit
these problems are more tractable since reactions are binary.  We also
notice that in templated ligation, the total cluster density,
$c=\sum_{k\geq 1}c_k$, and the monomer density $c_1$ satisfy
\begin{subequations}
\begin{align}
\label{c-lig}
\frac{dc}{dt} &= - \sum_{i, j\geq 1} c_i c_j c_{i+j} \,, \\
\label{c1-lig}
\frac{dc_1}{dt} &= -2c_1 \sum_{j\geq 1}c_j c_{j+1}\,.
\end{align}
\end{subequations}
and these cannot be solved recursively. In contrast, in aggregation
with mass-independent reaction rates, the analogous equations
$\frac{dc}{dt}=-c^2, ~\frac{dc_1}{dt}=-2cc_1$, are closed and readily
solvable.

Nevertheless, we can extract the essential long-time behavior from
Eqs.~\eqref{ck-lig} by invoking scaling.  Similarly to aggregation
\cite{Ernst85b,Ernst88} (see also~\cite{Leyvraz03,KRB} for reviews),
one expects that the mass distribution approaches the scaling form
\begin{equation}
\label{cc-Phi}
c_k(t) = c^2\Phi(k c)
\end{equation}
in the scaling limit $t\to\infty$, $k\to\infty$,  with $kc=$ finite.

The mass conservation statement \eqref{mass} and the definition of the
cluster density then lead to the integral constraints
\begin{equation}
\label{constraints}
\int_0^\infty dx\, x \Phi(x) = 1, \qquad \int_0^\infty dx\, \Phi(x) = 1\,.
\end{equation}
Here we have replaced the summations by integrations, which is
appropriate in the long-time limit where scaling is valid.

Substituting the scaling ansatz \eqref{cc-Phi} into \eqref{c-lig}, the
time dependence of the total cluster density is given byw
\begin{subequations}
\begin{align}
  \frac{dc}{dt} = - B c^4\,,
\end{align}
with
\begin{align}
\label{A-3}
B = \int_0^\infty dx \int_0^\infty dy\,\Phi(x)\Phi(y)\Phi(x+y)\,.
\end{align}
\end{subequations}
Solving the first of these equations gives the time dependence of the
cluster density
\begin{equation}
\label{c-lig-sol}
c \simeq (3 Bt)^{-1/3}\,.
\end{equation}
Using \eqref{c-lig-sol} and the scaling form \eqref{cc-Phi} we
conclude that the density of monomers is then
\begin{equation}
\label{c1-lig-sol}
c_1\simeq \Phi(0) (3 Bt)^{-2/3}\,.
\end{equation}

We can obtain an alternative expression for the amplitude $B$ from the
integral of the scaled mass distribution that is simpler than the the
double integral in Eq.~\eqref{A-3}.  Assuming $\Phi(0)>0$ and
substituting \eqref{c1-lig-sol} into \eqref{c1-lig} we obtain
\begin{equation}
\label{A-2}
B = \int_0^\infty dy\,\Phi^2(y)\,.
\end{equation}

To obtain the scaling function itself, we substitute the scaling form
\eqref{cc-Phi} into the governing equations \eqref{ck-lig} and find
that the scaling function $\Phi(x)$ obeys the non-linear
integro-differential equation
\begin{align}
  \label{ID}
B\left[2\Phi(x)+x\,\frac{d \Phi(x)}{dx}\right] &=2\Phi\int_0^\infty dz\,\Phi(z)\Phi(z+x)\nonumber \\
& - \Phi\int_0^x dy\, \Phi(y)\Phi(x-y)\,.
\end{align}
Notice that integrating \eqref{ID} over all $x$, we recover \eqref{A-3}.  In the limit $x\to 0$,
Eq.~\eqref{ID} reduces to \eqref{A-2}.  These relations serve as
useful consistency checks.

The time dependence given in Eq.~\eqref{c-lig-sol} together with the
equation \eqref{ID} for $\Phi$ constitutes a formal solution
cluster-mass distribution for the ligation reaction.  While the
explicit solution of \eqref{ID} is likely not possible, we have found,
in a direct way, the time dependence of the cluster densities.

\section{Summary and Discussion}

We introduced an aggregation model that is driven by templating.  Here
an aggregate of a specified mass acts as a scaffold upon which smaller
clusters meet and merge to create a cluster that also can act as a
scaffold.  Clusters whose mass is either larger than or equal to the
scaffold mass also undergo conventional aggregation.  Within the
mean-field description and also under the assumption that the all
reaction rates are mass independent, we solved for the kinetics of the
cluster mass distribution.

One basic result for this type of templating-controlled aggregation is
that the ensuing kinetics is much slower than in conventional
aggregation.  For the simple case where the scaffold is a mass-2 dimer
and two monomers must meet on this scaffold to create another dimer,
we found that the monomer density decays with time as $t^{-1/3}$,
while the densities of clusters of mass 2 or greater, as well as the
total cluster density, all decay as $t^{-2/3}$.  Therefore the decay
of the monomer density is {\em slower} than that of the cluster
density.  In conventional aggregation, the density of clusters of any
mass decays as $t^{-2}$, while the total cluster density decays as
$t^{-1}$.  Thus the monomer density decays {\em faster} than the
cluster density.  To summarize, the relation between the monomer and
cluster densities is $m\sim c^2$ in ordinary aggregation and
$m\sim \sqrt{c}$ in templating aggregation.

The templating reaction is three-body in nature, and this feature is
the underlying reason for the much slower kinetics compared to
conventional aggregation. Intriguingly, the relative cluster densities
$c_k/c$ in templating aggregation, Eqs.~\eqref{ratios}, are the same
as in ordinary aggregation driven by the source of small mass clusters
\cite{Leyvraz03, KRB}. This property is the chief qualitative
difference with ordinary aggregation where the mass distribution
approaches a scaling form in the long time limit.

We extended our model to a scaffold of arbitrary mass $L$, upon which
$L$ monomers meet and react to create another $L$-scaffold.  Another
natural extension that we studied is to allow multiple levels of
templating.  For a two-stage templating reaction in which dimers and
4-mers act as scaffolds to promote the reaction, we observed similar
behavior as in single-stage templating in which nearly all cluster
densities decay as $t^{-2/3}$.  In this two-stage reaction, it is only
the dimer density that now decays as $t^{-1/3}$.

We also introduced and investigated a templated ligation reaction,
where clusters of all masses serve as scaffolds, and there is no free
aggregation. The ensuing kinetics is much slower than in templating
aggregation. Namely, the total cluster density decays as $t^{-1/3}$,
while individual cluster densities decay as $t^{-2/3}$. In contrast to
templating aggregation, the mass distribution in templated ligation
approaches a scaling form.  However, even with mass-independent
ligation rates, templated ligation is theoretically more challenging
than templating aggregation.  We were able to only determine the time
dependence of basic observables, but the amplitudes of these decay
laws and the precise form of the scaled mass distribution remain
unknown.

At a theoretical level, templating-driven aggregation can be viewed as
a form of conventional binary aggregation, but with a non-trivial time
dependent source of scaffolds (either dimer or dimer and 4-mer) that
serve as the input to the aggregation process.  It is remarkable that
the non-trivial and slow time dependence of these small elemental
clusters modifies the densities of all heavier clusters so that the
overall aggregation reaction also has a slow time dependence compared
to conventional aggregation.  From the perspective of applications,
there are many situations where the notion of templating plays a major
role of many types of reactions.  In addition to the applications for
models of the origin of life mentioned in the
introduction~\cite{eigen1971selforganization,kauffman1971cellular,gilbert1986origin,Nowak08,Nowak09,Nigel15,Kaneko16,hordijk2019history,Steen14,Steen17,Kaneko18,tkachenko18,Kaneko23,SA21,Kaneko23},
other applications include, for example, self assembly of
colloids~\cite{yin2001template}, synthesis of exotic
materials~\cite{davis2001template,https://doi.org/10.1002/adfm.200300002,https://doi.org/10.1002/adma.201802349},
and protein aggregation~\cite{PhysRevLett.101.258101}.  Perhaps our
simple modeling can provide a starting point for understanding these
types of template-controlled reactions.

\bigskip\noindent We thank
Steen Rasmussen for stimulating conversations that helped nucleate
this project.  This work has been partially supported by the Santa Fe
Institute.

\bibliography{references-E}

\end{document}